\documentclass{aa}  

\usepackage{gensymb}
\usepackage{amsmath}
\usepackage{graphicx}
\usepackage{txfonts}

\usepackage{amsmath}
\usepackage{color}
\usepackage{graphicx}
\usepackage{graphbox}
\usepackage{natbib}
\usepackage{booktabs}
\usepackage{xspace}
\usepackage{array,multirow}
\usepackage{wrapfig}
\usepackage{hyperref}

\usepackage{xcolor} 

\begin{document}

\title{Evolution, speed, and precession of the parsec-scale jet \\ in the 3C 84 radio galaxy}

\subtitle{Super-resolved images from multi-epoch observations \\ at 43 GHz by the Very Long Baseline Array}

\author{M. Foschi \orcid{0000-0001-8147-4993}\inst{\ref{IAA}} \and 
        J. L. Gómez \orcid{0000-0003-4190-7613}\inst{\ref{IAA}} \and 
        A. Fuentes \orcid{0000-0002-8773-4933}\inst{\ref{IAA}} \and 
        I. Cho \orcid{0000-0001-6083-7521}\inst{\ref{KASI},\ref{Yonsei},\ref{IAA}} \and 
        A. P. Marscher \orcid{0000-0001-7396-3332}\inst{\ref{BU}} \and 
        S. Jorstad\orcid{0000-0001-6158-1708}\inst{\ref{BU}}}

\institute{
Instituto de Astrofísica de Andalucía (IAA-CSIC), Gta. de la Astronomía, s/n, 18008 Granada, Spain\label{IAA} \\ \email{foschimarianna@gmail.com}
\and Korea Astronomy and Space Science Institute, Daedeok-daero 776, Yuseong-gu, Daejeon 34055, Republic of Korea\label{KASI}
\and Department of Astronomy, Yonsei University, Yonsei-ro 50, Seodaemun-gu, Seoul 03722, Republic of Korea\label{Yonsei}
\and
Institute for Astrophysical Research, Boston University, 725 Commonwealth Ave., Boston, MA 02215, USA\label{BU}
}

\date{Received Month DD, 2024; accepted Month DD, 2024}

\abstract{We present high resolution images of the radio source 3C 84 at 43 GHz, from 121 observations conducted by the BEAM-ME monitoring program between 2010 and 2023. Imaging was performed using the recent forward modeling imaging method \texttt{eht-imaging}, which achieved a resolution of $80$ $\mu$as, a factor of $\sim$2-3 better than traditional imaging methods such as CLEAN. The sequence of images depicts the growth and expansion of the parsec-scale relativistic jet in 3C 84, clearly resolving a complex internal structure, showing bending in the jet, and changes in its launching direction and expansion speed. We report measurements of the expansion speed in time, which show that the jet undergoes three regimes, marked by the beginning and ending of a hot spot frustration phase. The images' high resolution allows us also to measure the projected launching direction as a function of time, finding an irregular variation pattern. Our results confirm previous studies of the morphological transition underwent by 3C 84 and provide quantitative measurements of the jet's kinematic properties over a decade time-scale.} 


\maketitle


\section{Introduction}

Some supermassive black holes at the center of galaxies generate collimated jets of ionized relativistic particles, which are accelerated by the strong magnetic fields surrounding the black hole and the accretion disk. These highly energetic and luminous jets propagate through the host galaxy and beyond, interacting with the the interstellar, intracluster and intergalactic media (ISM, ICM, IGM). Temperature, density and pressure differences between the plasma in the jet and the surrounding medium influence the jet's expansion, by shaping its shape and profile, and by affecting its direction and expansion speed.

The relativistic jet in the radio galaxy 3C 84 (NGC 1275) at the center of the Perseus Cluster, is a valuable source of information about the interactions between the ICM and the parse-scale jet.
3C 84 is a variable radio source \citep{Dent1966,Pauliny1966}, with bright X-ray emission \citep{Forman1972}. X-ray observations of the Perseus cluster by Chandra showed pairs of opposed bubbles in the intracluster medium, located at varying distances and directions with respect to the 3C 84 radio source \citep{Fabian2003}. A proposed explanation for these structures is that they are inflated by a precessing and restarting pair of jet and counter-jet \citep{Dunn2006}.
In the radio frequencies, 3C 84 has been observed since the 1950s by multiple very long baseline interferometry (VLBI) observations at both millimeter and centimeter wavelength.

\cite{Savolainen2023} provide an historical overview of radio observations of 3C 84, of which we give here a short summary. 
The source presents various lobe-like structures south and north of the core, from the parsec \citep[e.g.,][]{Walker2000,Asada2006} to the kiloparsec \citep{Pedlar1990} scales, which may indicate a repeatedly restarting jet.
Observations at the parsec scale from the 2010s have shown the presence of a dim radio lobe, named C2 by \citet{Nagai2010}, and a bright radio lobe (C3), that is connected to the core via a limb-brightened structure \citep{Nagai2014} (see the top-left panel in Fig. \ref{fig:ehtimclean}). The times at which these components were emitted coincide with periods of increased brightness, in the early 1960s for C2 \citep{Nesterov1995} and in the early 2000s for C3 \citep{Nagai2010}.
In more recent years, higher resolution observations have been able to resolve the internal structure of the parsec-scale jet in 3C 84. \citet{Giovannini2018} presented results at 22 GHz with a global array of ground antennas plus the space antenna RadioAstron \citep{Kardashev2013}. The reconstructed image clearly shows strong limb-brightening and a wide opening angle near the core, followed by a quasi-cylindrical jet profile. \citet{Giovannini2018} suggested that the cylindrical profile may be due to the jet connecting C1 to C3 being embedded in a uniform-pressure cavity carved by past activity of the jet. This is supported by other observations by RadioAstron at 5 GHz \citep{Savolainen2023}, which show that the C2 and C3 components are both surrounded by low-intensity emission from a cocoon-like structure. \cite{Savolainen2023} explain that C3 is the head of a restarted jet that is interacting with the ISM. In the interaction, energy is transferred to the ISM, heating the gas that forms the cocoon. They also suggest that the cocoon-like structure could be caused by the jet moving trough a multi-phase medium consisting of gas clouds of different sizes and densities. Embedding in a clumpy medium is supported by results from \citet{Nagai2017} and \citet{Kino2018,Kino2021}. In particular, \citet{Kino2021} analyzes 43 GHz images of 3C 84 from the Very Long Baseline Array (VLBA) from 2012 to 2020. They track the motion of a hot spot and, in 2016-2017, they observe a year-long frustration phase, during which the hot spot follows a circular trajectory after reaching the edge of C3, instead of propagating further through the jet. They attribute this event to a collision between the head of the jet and a compact dense cloud. After the collision, the jet breaks through the cloud, deviating its expansion direction to the west and transitioning from an FR II- to FR I-class radio lobe morphology.
However, despite the significant amount of studies, observations of 3C 84 have provided either repeated images of the jet in low resolution \citep{Kino2018,Kino2021} or hard-to-repeat single-epoch images in high-resolution \citep{Giovannini2018}. This hinders a proper study of the kinematics of the plasma in the jets and the dynamics of the jet expansion. 

In this work we present a re-imaging of all 121 VLBA observations of 3C 84 at 43 GHz, obtained by the BEAM-ME monitoring program \citep{Jorstad2016} from 2010 to 2023, using the Regularized Maximum Likelihood (RML) imaging method \texttt{eht-imaging} \citep{Chael2018}. With \texttt{eht-imaging}, we obtained images of the parsec-scale jet at a resolution of $\sim$80 $\mu$as, which is  $\sim$2-3 times higher than the nominal beam used to convolve the CLEAN images, whose average among different epochs is (280, 150) $\mu$as. RML methods produce super-resolved images by incorporating reasonable prior assumptions that regularize the image. These methods have proven to achieve higher fidelity at super-resolution than CLEAN \citep[see e.g.][]{Fuentes2023}. Thanks to the combination of the super-resolving power of RML methods and the constant monitoring by the BEAM-ME program, we were able, for the first time, to observe the evolution of the overall and internal structure dynamics of the parsec-scale jet, over a 12 year period. The images we are presenting resolve the hot spots and the internal structure of the jet, as well as the connection between the limb-brightened structure and the core. 
With this resolution, it is also possible to resolve the front of the jet head, discerning the expansion of the jet from the motion of components through the jet. In this work we considered the source redshift $z=0.0176$ \citep{Strauss1992}. In continuity with previous publications on 3C 84, we assumed a $\Lambda$ cold dark matter cosmology with $H_0=70.7$ km s$^{-1}$ Mpc$^{-1}$, $\Omega_M=0.27$, and $\Omega_\Lambda=0.73$, resulting in 1 mas in the image plane corresponding to 0.35 pc. 
 
The layout of the paper is as follows. In Sect. \ref{sec:dataimaging} we give details of the observed data and explain the method used to image them. In Sect. \ref{sec:results} we present the imaging results, provide a quantitative estimate of the jet's speed and direction and discuss the evolution of the jet in the context of previous observations of the source. We summarize and discuss our results in Sect. \ref{sec:conclusions}.


\section{Data and Imaging}\label{sec:dataimaging}


\subsection{Observations}

We analyzed data from the BEAM-ME monitoring program conducted by Boston University (previously named VLBA-BU-BLAZAR)\footnote{\texttt{https://www.bu.edu/blazars/BEAM-ME.html}}, which observes multiple gamma-ray blazar and radio sources using the VLBA at 43 and 86 GHz \citep{Jorstad2016}. We focused on total intensity observations of the radio source 3C 84 in the Perseus cluster, at 43 GHz, conducted on a roughly monthly basis from late 2010 until early 2023, resulting in a total of 121 individual epochs.

The BEAM-ME data is already fully calibrated and self-calibrated to the CLEAN images provided in the archive. However, in order to avoid any bias from possible residual calibration errors or from the self-calibration, we chose to run the first imaging step using only closure quantities. The archival data was already time averaged with a 30s interval, so no additional time averaging was performed before imaging.

\subsection{Imaging procedure}

The data was imaged using \texttt{eht-imaging}, a forward modeling imaging method for VLBI observations \citep{Chael2018}. This method defines the image as a discrete square matrix of flux density values, $\textbf{I} = \{I_{ij}\}$, and optimizes these values to minimize the objective function:
\begin{equation}
    J(\textbf{I}) = \sum_\text{D} \alpha_D \chi^2_D(\textbf{I},\textbf{d}) - \sum_\text{R} \beta_R S_R(\textbf{I})
\end{equation}
where the first sum runs over the reduced $\chi^2$ of different data products $D$ computed from the image $\textbf{I}$ and the observed data $\textbf{d}$, while the second sum runs over various regularizers $R$ that impose additional correlations among pixel values, thus constraining the possible solutions to the ill-posed problem of VLBI imaging \citep{eht2019m87iv}. The coefficients $\alpha_D$ and $\beta_R$ that weight the data terms and the regularizers are hyperparameters of the method.

The imaging procedure followed these steps:
\begin{enumerate}
    \item Iteratively apply \texttt{eht-imaging}'s optimization step, using only log closure amplitudes and closure phases as data products in the objective function. Stop the optimization when the reduced $\chi^2$ of both data products decreases by less than 2\% in one step. 
    \item Perform self-calibration of amplitudes and phases to the obtained image, in order to correct for potential residual station-based errors.
    \item Re-apply \texttt{eht-imaging}'s optimization steps, this time using both complex visibilities and closure quantities as data products in the objective function. Stop the optimization when the $\chi^2$ of both data products decreases by less than 1\%.
\end{enumerate}
The \texttt{eht-imaging}'s optimization step mentioned in point 1 and 3 consists of alternating between a series of quasi-Newton gradient descent steps and blurring of the resulting image with a Gaussian kernel with FWHM equal to 1.5 times the nominal resolution of the array ($\sim$150 $\mu$as). The blurring step prevents the optimizer from becoming trapped in the local minima of the objective function.
The images obtained with this procedure provide a good fit to the data, with an average reduced visibility $\chi^2$ of 1.61. We note that, in most cases, the images obtained after step 1 are very similar to those obtained after step 3, meaning that the closure quantities are sufficient to constrain the images and that self-calibration only provides small refinements to the final images. The gains obtained from the self-calibration step were negligible, which was expected since the archival data were already self-calibrated using CLEAN imaging.

\paragraph{Field of view} The chosen field of view was increased linearly from 5.1 mas for the earliest epoch to 13.2 mas for the latest one, following the growth of the emitting region in the jet, as seen in the archival CLEAN images. Accordingly, the number of pixels was increased from 170 to 440, while the pixels' size was kept constant at 30 $\mu$as\footnote{The images presented in the paper are cropped with respect to the original field of view, to be displayed in the most effective way.}.

\paragraph{Initialization image} The initialization image for the optimization process was chosen to be an elliptical Gaussian with major axis rotated 6$\degree$ clockwise from the North, with (FWHM$_x$, FMHM$_y$) values ranging linearly from (385, 1375) $\mu$as to (910, 3250) $\mu$as, so to match the average direction and increasing dimension of the jet.
For a few epochs (specifically, from 30/04/2022 to 06/12/2022), the optimizer would not converge if initialized to a Gaussian. We attribute this to the combination of a suboptimal coverage and a complex jet morphology, which could not be well approximated by a Gaussian.
In these cases, we initialized the optimizer to the CLEAN image provided by the BEAM-ME program, blurred with a Gaussian kernel of FWHM equal to twice the nominal resolution of the array. We tested the effect of initializing with the CLEAN image, instead of a Gaussian, on the other epochs and found that the final image was not noticeably affected by the prior image choice. 

\paragraph{Regularizers} We made minimal use of the regularizers, since the uv-coverage of the array is sufficiently dense. In step 1 of the imaging procedure, the total flux of the image cannot be constrained by closure amplitudes, so we used a flux regularizer to constrain it to the maximum amplitude of the shortest baseline. In both step 1 and step 2, we used the entropy regularizer to constrain the emission in the center of the image and the $\ell_1$ regularizer to encourage sparsity, since significant portions of the images were expected to have no emission. The exact regularizers' weights, along with other imaging parameters, are reported in Table \ref{tab:hpar}.

\begin{figure*}
    \resizebox{\hsize}{!}{
        \includegraphics[width=\hsize]{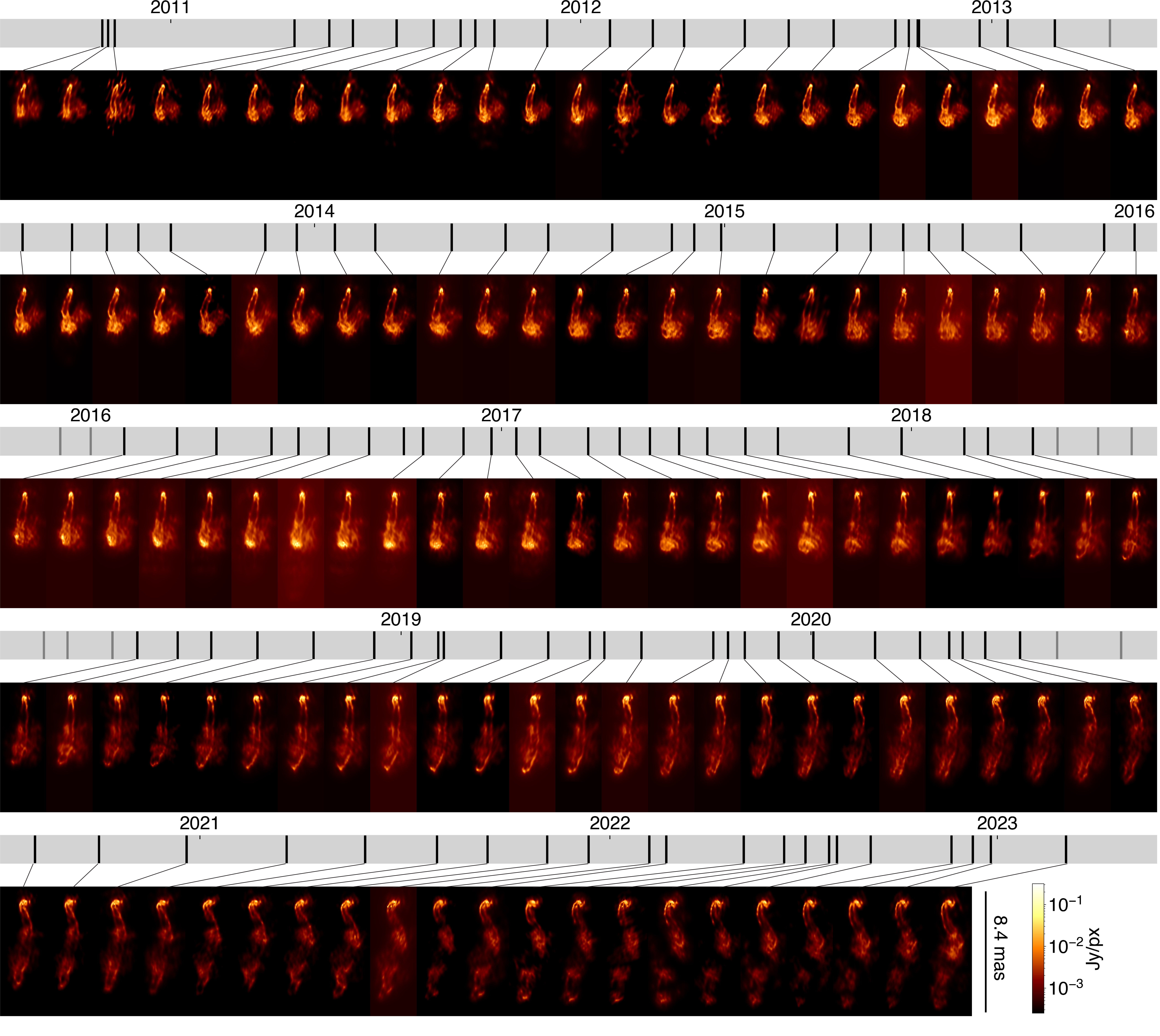}}
    \caption{Radio source 3C 84 as observed from 2010 to 2023 by the VLBA at 43 GHz and imaged with \texttt{eht-imaging}. The black vertical lines mark the dates of the observation epoch corresponding to each image. The time evolution of the source is shown in the online movie.}
    \label{fig:timeline}
\end{figure*}

\begin{figure}
    \centering
    \includegraphics[width=\hsize]{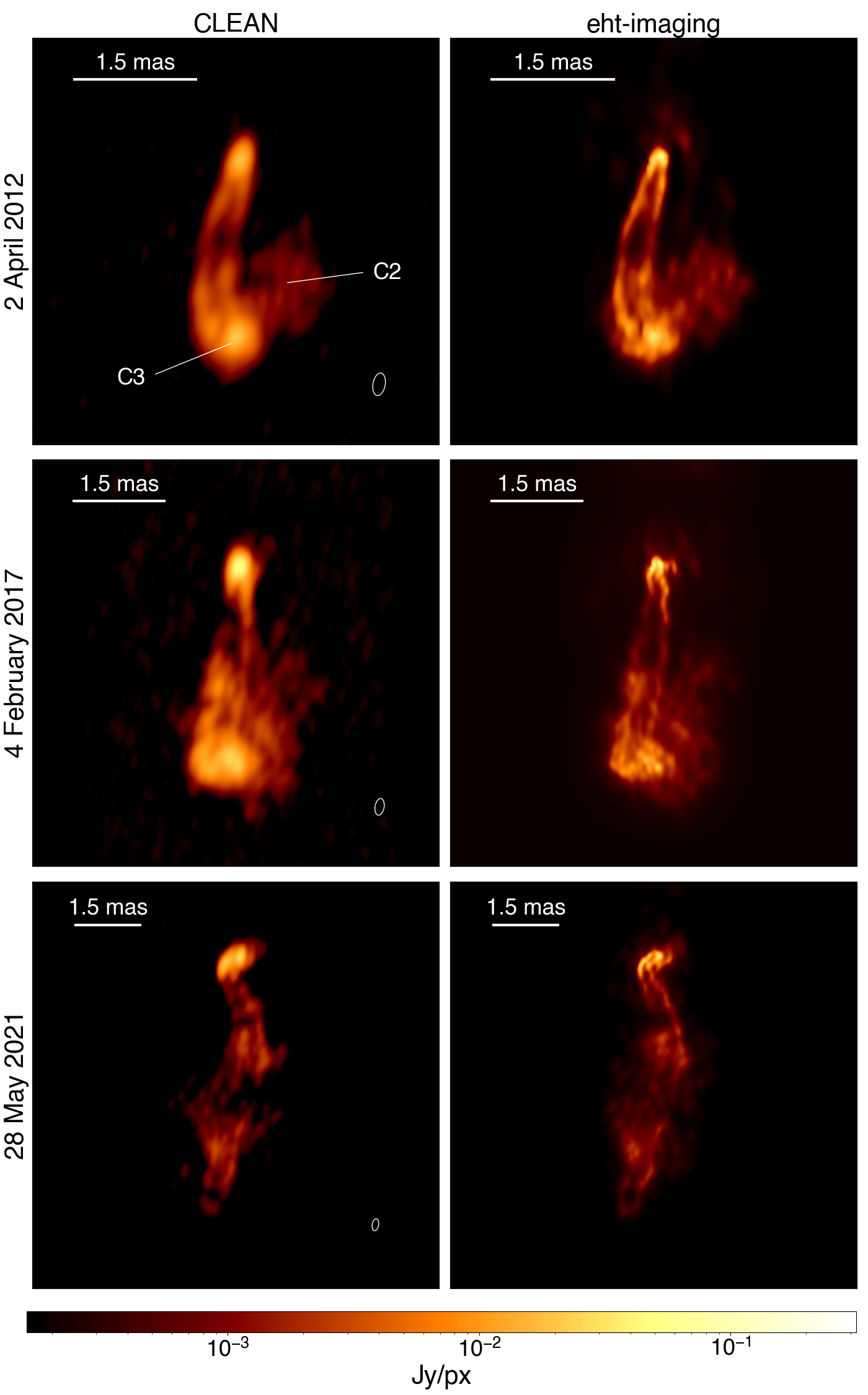}
    \caption{Comparison between images of radio source 3C 84 from VLBA observations at 43 GHz, obtained with the CLEAN (left) and \texttt{eht-imaging} (right) imaging methods.}
    \label{fig:ehtimclean}
\end{figure}

\paragraph{Images} Imaging results from all epochs are presented in Fig. \ref{fig:timeline} and in the corresponding online movie, while Fig. \ref{fig:ehtimclean} shows the comparison between CLEAN and \texttt{eht-imaging} images for a few selected epochs. The \texttt{eht-imaging} images we obtained are consistent with the CLEAN images provided by the BEAM-ME program, but in higher resolution.
This enables a more precise detection of the outlines of the bright limbs and the edge of the jet head, as well as the resolution of intra-jet features and the jet orientation at the sub-parsec scale.

\begin{table}[htbp]
    \caption{Imaging hyperparameters used in the \texttt{eht-imaging} pipeline.} 
    \centering
    \begin{tabular}{ll}
    \hline\hline
    \textbf{Image parameters} & \textbf{Values} \\ 
    pixel size & 30 $\mu$as \\
    number of pixels & 170 - 440 \\
    field of view & 5.1 - 13.2 mas \\
    \hline
    \textbf{Observation parameters} &  \\ 
    added systematic error & 0.5 \% (of visibility amplitudes) \\
    time averaging & 30 s \\
    \hline
    \textbf{Prior parameters} &  \\ 
    prior image & Gaussian / CLEAN image\\
    Gaussian fwhm (maj axis) & 1375 - 3250 $\mu$as \\
    Gaussian fwhm (min axis) & 385 - 910 $\mu$as \\ 
    Gaussian orientation & -6$\degree$ \\
    \hline
    \textbf{Optimization parameters} &  \\ 
    max N iterations & 200\\
    \textbf{data term weights  (steps 1 - 2)} & \\
    vis     & 0 - 1 \\
    amp     & 0 - 1 \\ 
    cphase  & 1 - 1 \\
    logcamp & 1 - 1 \\
    \textbf{regularizer weights (steps 1 - 2)} & \\
    flux     & 50 - 10  \\
    entropy  & 0.1 - 1  \\
    $\ell_1$ & 0.1 - 10 \\
    \hline
    \label{tab:hpar}
    \end{tabular}
\end{table}

\section{Jet analysis}\label{sec:results}

The sequence of jet images in Fig. \ref{fig:timeline}, reveals various features of the jet evolution. 
In the first images (corresponding to epochs in late 2010), the jet presents clear limb-brightening, with the two limbs originating from an unresolved core and undergoing a slight counter-clockwise bending. A diffuse emission feature is evident in the bottom-right part of the jet. Over the years, the limb-brightening remains unchanged as the jet increases in length, reaching $\sim$4 times its initial size. 
The jet launching direction changes over the years, slowly rotating counterclockwise from 2011 to 2015, slowly rotating back to the initial direction from 2015 to 2016, maintaining the same direction during 2017-2018, then drastically rotating clockwise from 2019 to 2022.
The limb-brightened structure extends until close to the core, to which it connects at a high opening angle, confirming the analysis from \citet{Savolainen2023}. From 2019, the core undergoes a severe twisting due to the clockwise rotation, which complicates its structure, while a secondary component appears in 2017 west to the core and persists until the end of the considered time window. This was also observed by \citet{Punsly2021}, who modeled the complex core structure with two and three components aligned in the east-west direction.

Various components can be tracked moving along the jet, notably a bright spot is seen approaching the head of the jet, "bouncing" against it and then dissipating, in the time span between late 2015 and early 2017. Our results confirm the hot spot's counterclockwise trajectory reported by \citet{Kino2021}. However, our images show that the hot spot appears at the end of 2015 and dissipates at the beginning of 2018 as the jet pierces trough the lobe. The 2015 hot spot's flip and the 2018 hot spot's breakout reported by \citet{Kino2021}, should be attributed to a component mismatch caused by insufficient resolution.

At the end of 2010, the jet presents a straight morphology. However, a lobe begins to form in late 2011 and undergoes a significant expansion from early 2013 to early 2017. The inflation coincides with a slowdown in the jet's expansion velocity, followed, in 2018, by a burst through the inflated bubble and an increase in the expansion speed.
This confirms the abrupt morphological transition from a FR II- to FR I-class radio lobe observed by \cite{Kino2021}, to which we add a gradual opposite transition from FR I to FR II observed from 2010 to 2013.
From late 2020 to the end of the considered epochs, some portions of the jet appear darkened. This could be caused by a lower local emissivity, a change in the viewing angle or by the presence of an absorbing foreground. With respect to this last hypothesis, it is to be noted that 3C 84 is likely surrounded by an accretion disk associated with ionized gas, which absorbs and obscures the inner section of the counter-jet \citep{Walker2000,Fujita2017}.

\subsection{Feature extraction}

Traditionally, VLBI images of relativistic jets have been analyzed by fitting Gaussian components to the bright features of the jet and tracking the motion of these components over time, which is known as "model fitting". 
This was the best approach to analyze jet dynamics when the resolution was not sufficient to resolve features inside the jets. However, new super-resolution imaging methods such as \texttt{eht-imaging} allow us now to resolve intra-jet features \citep[see][]{Janssen2021,Fuentes2023,Savolainen2023,Park2024}, making it unnecessary to approximate and oversimplify jet images using a set of Gaussians. 
In these cases, model fitting is not an adequate tool. Instead, case-by-case methods should be chosen, depending on the features visible in the image.
Because of the winding and evolving jet structure observed in our images of 3C 84, we characterized the jet by measuring its maximum radial expansion from the core (see Sect. \ref{sec:jetlength}) and by tracing the profiles of the two bright limbs, from which we computed the overall jet outline and the jet launching direction (see Sect. \ref{sec:jetdirect}).

\subsection{Alignment}

To compare features extracted from images at different epochs, it is crucial that the images are properly aligned. To align the images with respect to the jet core, we applied the following steps:
\begin{enumerate}
    \item Locate the brightest pixel in each image.
    \item Apply a circular mask of 1 mas radius, centered in the brightest point, setting to zero every pixel outside the mask.
    \item Shift each image by the amount that maximizes the cross-correlation between the masked image and the masked image of the previous epoch.
\end{enumerate}
This procedure was effective for correctly aligning all epochs except those between 03/11/2019 and 06/11/2021. For those epochs, a simple cross correlation alignment was not effective because of the rapid twisting of the core region, so an additional shift, linearly spaced from 35 $\mu$as up to 246 $\mu$as, was applied after the cross-correlation shift.

\subsection{Edge fitting}

\begin{figure}
    \centering
    \includegraphics[width=\hsize]{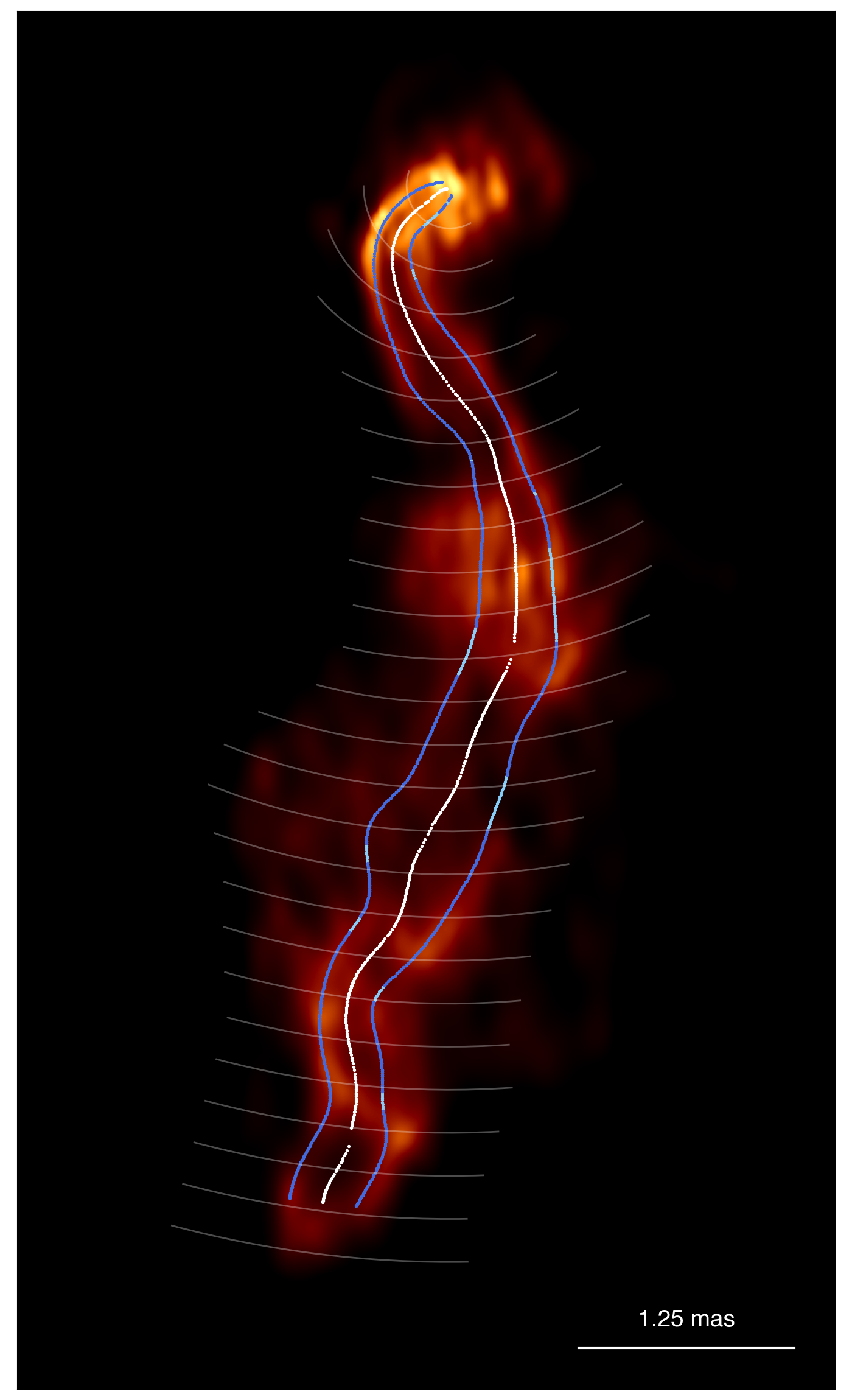}
    \caption{Example of the fitted edges and jet direction for the jet in epoch 21-07-2021. Dark blue points represent fitted edge points, while light blue points represent interpolated edge points. White points represent the jet outline. A subset of the circular sections used to detect the edge points is shown as thin white lines.}
    \label{fig:edges}
\end{figure}

The most straightforward way to characterize the features present in the jet is to outline its edges by tracking the position of the two bright limbs. To this purpose, as shown in Fig. \ref{fig:edges}, we considered circular sections of the image, centered around the jet core. We find the two highest emission peaks along each of these profiles and assign their positions to the two bright edges, creating two set of points outlining each edge (blue points in Fig. \ref{fig:edges}). In some sections, the position of one or both limbs could not be detected using this method. In such cases, the limb position was determined by interpolating between the positions in the preceding and following sections. The overall jet outline (white points in Fig. \ref{fig:edges}) was determined by the set of midpoints of the distance segments between each point of one limb and the other limb. 

\subsection{Jet launching direction}
\label{sec:jetdirect}

\begin{figure*}[h!]
    \resizebox{\hsize}{!}{
        \includegraphics[width=\hsize]{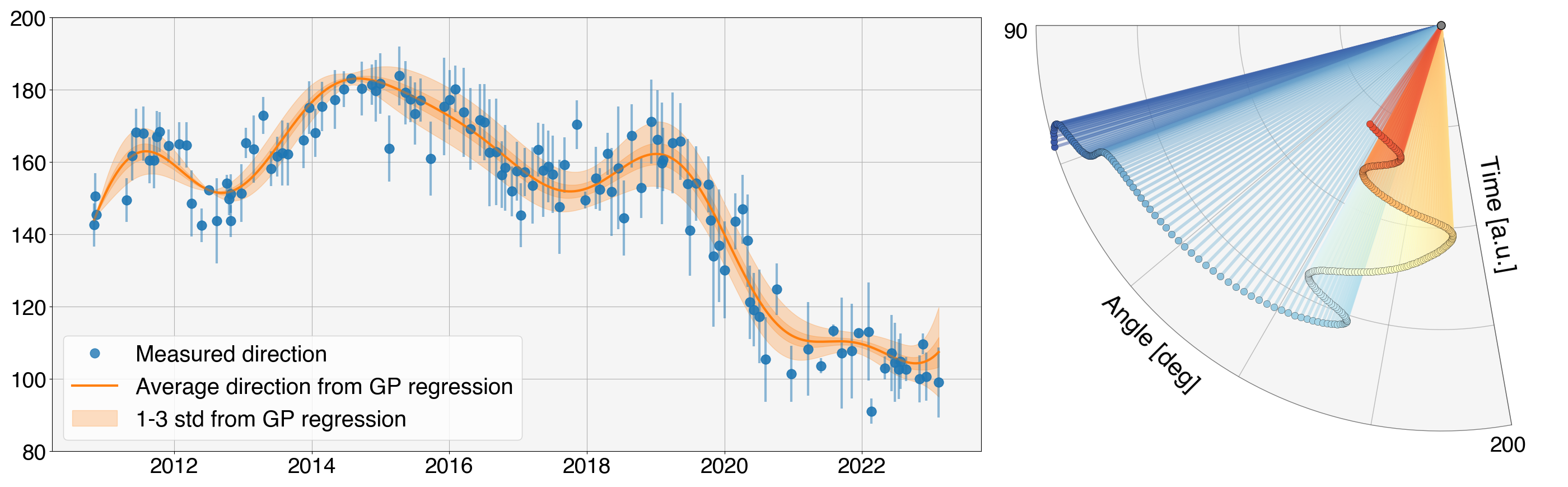}}
    \caption{(Left) Angle of the jet launching direction as a function of time. The angle is measured East of North. In blue, the measured direction from each epoch, in orange, the average of a GP regression to the data. The orange shading indicates the 1 sigma (dark color) and 3 sigma (light color) uncertainty from the GP regression. (Right) Average direction from the GP regression, plotted in angular coordinates for a more intuitive representation of the direction change. Red colors correspond to earlier epochs while blue colors correspond to later epochs.}
    \label{fig:jetdirect}
\end{figure*}

The local jet direction at each section of the jet is defined by the vector tangent to the jet outline curve. We determined the initial jet launching direction by averaging the tangent vectors corresponding to the segments of the jet outline within 90 $\mu$as from the core, to avoid resolution-induced biases. The standard deviation associated to the average was assigned as the error on the measured direction.
The left panel in Fig. \ref{fig:jetdirect} shows the angle corresponding to the jet launching direction as a function of time. The points in blue represent the measured direction from each epoch, while the orange lines indicate the average and standard deviation of a Gaussian Process (GP) regression to the data.
The right panel in Fig. \ref{fig:jetdirect} shows the average jet launching direction from the GP regression with the angle plotted in angular coordinates for a more intuitive representation of the directional shift.
The plot shows that the orientation of the jet within the first 0.03 pc from the core undergoes several irregular oscillations. We observe a 20$\degree$ oscillation from the beginning of 2011 to the end of 2012, a 40$\degree$ oscillation from early 2013 to early 2017, a 10$\degree$ shift until early 2018, followed by a clear 60$\degree$ shift until 2021 and a constant trend in late 2021 and 2022. The overall change in the jet's orientation spans 80 degrees.

A change in the direction of the jet has been proposed by \citet{Dunn2006} to explain the presence of X-ray holes at different orientations with respect to the core on the kilo-parsec scale. They suggest two possible causes for jet precession: a binary black hole system would make the jet of the primary black hole undergo a regular precession \citep{Katz1997}, while an instability in the accretion disk or a misalignment between the black hole spin and the accretion disk axis could cause the disk to warp, resulting in a stochastic jet precession \citep{Pringle1997}. 
However, they estimate a precession time scale of the order of $10^7$ years. Here instead we observe a drastic irregular variation in the jet's direction over a time scale of a few years. It is possible that the jet in 3C 84 undergoes precession cycles over different time scales.

\subsection{Jet length and expansion speed}
\label{sec:jetlength}
To measure the jet length, we first computed the longitudinal intensity profile of the jet by identifying the highest intensity value of the image along circular sections centered on the core. We defined the jet length, as projected in the image plane, as the maximum distance from the core reached by the head of the jet.
The head of the jet was identified as the point where the longitudinal jet profile dropped below the average noise floor, which was computed as the mean image value in a portion of the image not covered by the jet. The threshold value was adjusted for epochs from June 2016 to January 2017 to accommodate for a significantly higher noise floor and for epochs from May to September 2021 to take into account the darkening of portions of the jet. The darkening appears in various portions of the jet in later epochs, but it affects the measurement of the jet length only in mid 2021 because a darkened portion coincides with the jet head, as shown, for example, in the lower panel of Fig. \ref{fig:ehtimclean} or in Fig. \ref{fig:edges}.
The uncertainty associated with the jet length measurements was taken as the pixel size used in the imaging process (30 $\mu$as), under the assumption that the limited resolution is the main source of uncertainty for this measurement.

\begin{figure}[thbp]
    \centering
    \includegraphics[width=\hsize]{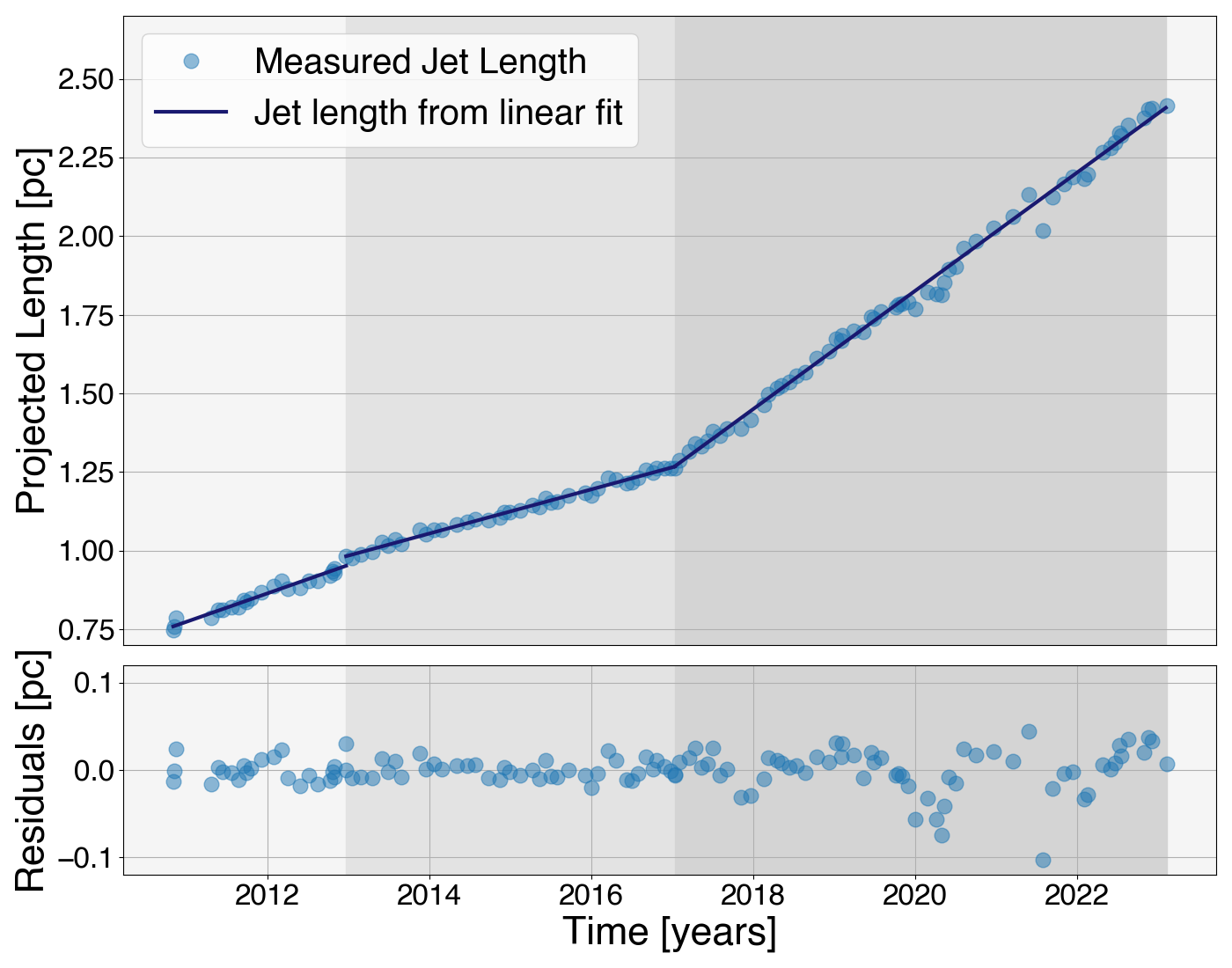}
    \caption{(Top) Measured jet length as a function of time (points) and piece-wise linear fit (continuous line). Uncertainties on the measurements are not shown because the error bars are smaller than the size of the points. (Bottom) Residuals of the linear fits.}.
    \label{fig:jetlen}
\end{figure}

Figure \ref{fig:jetlen} shows the jet length as a function of time. As visible in the online movie, three different trends are evident. Until the end of 2012 the jet length increases linearly, from the beginning of 2013 to the beginning of 2017 the increase occurs at a lower rate, and finally, from 2017, the expansion occurs at a higher speed than the initial one. A possible explanation for the speed change, is that the jet propagates across a medium with different densities, possibly shaped by past activity of the jet. For each of these three expansion regimes, we performed a linear fit to the jet length, to compute the expansion velocity. The residuals of the linear fits are shown in the lower panel of Fig. \ref{fig:jetlen} and do not show significant trends, meaning that the expansion in each regime was indeed proceeding at a constant speed.
From the measured velocities projected in the image plane, we computed the true de-projected velocities of the jet front, accounting for special relativistic effects and assuming an inclination of $\theta=18\degree$ \citep{Tavecchio2014}.

Table \ref{tab:jetv} reports the values of the apparent speed in the image plane and the de-projected physical speed. The reduced $\chi^2$s of the linear fits are also reported in Table \ref{tab:jetv} and show that the uncertainties are properly estimated for the first two regimes, while they might be slightly underestimated in the last case. This might be due to some portions of the jet being obscured in the latest epochs, which causes a higher uncertainty in the detection of the jet head.
Previous estimates of the speed of the 3C 84 jet head, in our considered time period, measured an average apparent speed of $0.27\pm0.02$ c between 2007 and 2013 \citep{Hiura2018}, and an average apparent speed of $0.33$ between 2003 and 2020 \citep{Kino2021}. \citet{Weaver2022} measured the apparent speed of the components in 3C 84 in the period from 2010 to 2019, through model fitting. The speed of the components is not necessarily the same speed of the jet expansion, but it can be useful to compare the two. In the period between 2010 and 2012, the apparent speed of components C2, C3, and C6 are 0.27 c, 0.21 c, 0.34 c, matching our estimate of the expansion speed of 0.29 c. In the period between 2013 and 2015, \citet{Weaver2022} reported a speed of 0.84 c for component C9, matching the observation of the hot spot frustration, which moves at a higher speed than the jet expansion. Finally, from 2016 to 2019, they report component C10 moving at a speed of 1.38 c, indicating a drastic increase in velocity, which matches our measured expansion speed increase.

\begin{table*}[bhtp]
    \caption{Jet front expansion velocity computed from the linear fit for each of the three expansion regimes. We report both apparent velocities projected in the image plane and the corresponding physical velocities, assuming an inclination angle of $\theta=18\degree$.} 
    \centering
    \begin{tabular}{lllll}
    \hline\hline
    time period & apparent speed & apparent speed & physical speed  & $\chi^2$\\ 
     & ($\mu$as/day) & (c) & (c)  & \\ 
    \hline
    01/11/2010 - 21/12/2012 & $0.71\pm0.03$   & $0.29\pm0.01$&   $0.50\pm0.02$   & 1.7\\ 
    21/12/2012 - 14/01/2017 & $0.549\pm0.009$ & $0.228\pm0.004$& $0.434\pm0.007$ & 0.81\\
    14/01/2017 - 11/02/2023 & $1.47\pm0.02$   & $0.61\pm0.01$&   $0.69\pm0.01$   & 7.0\\
    \hline
    \label{tab:jetv}
    \end{tabular}
\end{table*}

\section{Conclusions}\label{sec:conclusions}
In this work, we presented images of the parsec-scale jet in 3C 84 from 121 VLBA observations at 43 GHz, from late 2010 to early 2023. Thanks to the super-resolution enabled by the RML imaging method \texttt{eht-imaging}, our images resolve the internal structure of the jet, its profile and edges and different hot spot components moving inside the jet. We conducted a quantitative kinematic analysis of the jet's expansion and precession.

We observed the expansion of the limb-brightened restarted jet, and irregular oscillations of its launching direction. The limb brightening suggests a spine-sheath structure, where the inner part of the jet moves at a higher speed than the outer part, and could explain the bright gamma-ray emission observed by \citet{Abdo2009}.
The parsec-scale jet gradually transitioned from a FR I- to an FR II-like morphology from 2010 to 2013, while in 2017 it abruptly transitioned back to an FR I-like morphology. After the second transition, portions of the further side of the jet appear darkened.
We confirmed previous observations of a hot spot frustration during the epochs preceding the jet breakthrough. The hot spot is observed following a counterclockwise trajectory around the lobe of the jet head. However, contrary to prior observations, we distinguished the frustrated hot spot from other components present before 2016 and after 2017.
We measured the jet direction within the first 0.03 pc from the core and observed it undergoing various irregular oscillations, spanning an overall angle of 80$\degree$.
We measured the expansion speed of the jet head (C3 component) and observed three separate regimes of linear expansion. The first regime corresponds to the FR I - FR II transition, the second one to the inflation of the jet head in the FR II state and the third one to the expansion following the FR II - FR I transition. We measure apparent speeds of 0.29$\pm$0.01 c, 0.228$\pm$0.004 c, and 0.61$\pm$0.01 c, respectively.

Overall our results indicate that the jet is propagating in an irregular ISM, characterized by the presence of clumps of denser material, that affects the jet's speed, direction, and morphology. The presence of localized absorbing gas in front of the jet may also be an explanation for its local darkening, which alternatively may be due to an increase of the viewing angle or a change in emissivity. The observed evolution of the parsec-scale jet suggests that the presence of radio lobes may be a temporary stage in the evolution of a jet, caused by density differences in the ambient medium. In 3C 84 this seems to be confirmed by the presence of ancient lobes 
at multiple scales.
A possible reason for the irregular jet precession can be an instability in the disk or a misalignment between the angular momentum of the accretion disk and the spin of the black hole \citep{Dunn2006,Pringle1997}, which may cause a warping of the disk and a stochastic variation of the jet's direction.

Our results show that by using innovative super-resolving imaging methods it is possible to resolve complex features in the jet structure, which were previously accessible only at higher observing frequencies or with significantly longer baselines.
This marks a change in the way that jet features can be analyzed, shifting from the fitting of simple Gaussian components to more specific analysis adapted to the jet morphology.
For example, similarly to \citet{Park2024}, our \texttt{eht-imaging} images clearly highlight the limb-brightening structure, which allowed us to trace the jet edges and precisely track the changes of the jet direction in time. Future imaging of jet sources with new imaging methods could uncover additional cases of limb-brightening that were previously undetectable due to insufficient resolution.

\begin{acknowledgements}
    We thank Marie-Lou Gendron-Marsolais (Université Laval and IAA-CSIC) for useful discussions on this work.
    The project that gave rise to these results received the support of a fellowship from "la Caixa" Foundation (ID 100010434), with fellowship code LCF/BQ/DI22/11940027. 
    Authors M. Foschi, J. L. Gómez and A. Fuentes acknowledge financial support from the Severo Ochoa grant CEX2021-001131-S funded by MCIN/AEI/ 10.13039/501100011033.
    The work at the IAA-CSIC was supported in part by the Spanish Ministerio de Economía y Competitividad (grant number PID2022-140888NB-C21). 
    I.C. is supported by the KASI-Yonsei Postdoctoral Fellowship program.
    This study makes use of VLBA data from the VLBA-BU Blazar Monitoring Program (BEAM-ME and VLBA-BU-BLAZAR; \texttt{http://www.bu.edu/blazars/BEAM-ME.html}), 
    funded by NASA through the Fermi Guest Investigator grants, the latest is 80NSSC23K1508. The VLBA is an instrument of the National Radio Astronomy Observatory. The National Radio Astronomy Observatory is a facility of the National Science Foundation operated by Associated Universities, Inc.
\end{acknowledgements}

\bibliographystyle{aa}
\bibliography{main.bib}

\end{document}